\documentclass[%
twocolumn, superscriptaddress,
nofootinbib,
 amsmath,amssymb,
 aps,
pre,
]{revtex4-1}
\usepackage[all]{xy}
\usepackage{mathrsfs}
\usepackage{graphicx}
\usepackage{epstopdf}
\usepackage{dcolumn}
\usepackage{bm}
\usepackage{soul}
\usepackage{color} 
\usepackage{CJK}
\usepackage{ctable}  
\usepackage{hyperref}
\usepackage{dsfont}
\usepackage{booktabs} 

\definecolor{myorange}{RGB}{255,117,40}

\begin{document}
\title{Energetically efficient, mediated mechanical system for precise control of hoisting operations}
\author{Mikel Palmero}
\email{mikel.palmero@ehu.eus}
\affiliation{Applied Physics, University of the Basque Country (UPV/EHU)}
\affiliation{EHU Quantum Center, University of the Basque Country, UPV/EHU}
\author{Juan Gonzalo Muga}
\affiliation{Physical Chemistry, University of the Basque Country (UPV/EHU)}
\affiliation{EHU Quantum Center, University of the Basque Country, UPV/EHU}
\author{Ander Tobalina}
\affiliation{Applied Mathematics, University of the Basque Country (UPV/EHU)}
\affiliation{EHU Quantum Center, University of the Basque Country, UPV/EHU}
\begin{abstract}
%
  We introduce a mechanical control system for energy efficient and robust hoisting crane operations.
  The control system efficiently translates the harmonic motion 
  of a spring loaded mediating system into the desired driving of the load, recyling most of the employed energy for subsequent operations. The control output is a shortcut-to-adiabaticity protocol borrowed from quantum mechanics.
  The control system reduces the single operation consuption in realistic working regimes, but it is in cyclical processes where the energetical advantage becomes substantial. 
  The design of the control system and the control output is flexible enough to 
  allow additional optimization of the robustness against perturbations. 
\end{abstract}
\maketitle

\paragraph*{\textbf{Introduction}}

Hoisting plants are essential to a variety of economic activities. 
They are intensively used in harbors, factories, logistic centers and construction sites among many others. 
They are typically powered either by diesel or by electrical energy, 
which remains mainly produced by burning fossil fuels \cite{IEA2023}. 
Hoisting operations have increasingly been approached from an energy efficiency perspective \cite{PATTERSON1996}, leading to control schemes devised to reduce energy consumption. A possible approach aims to recover the energy outflow during lowering operations using supercapacitors \cite{Iannuzzi2009} or flywheels \cite{Ahamad2019}. Alternatively, specific evolutions of crane parameters have been designed based in control theory \cite{burns2001}. For instance, optimal control has been used to reduce the energetic bill in various processes of crane operation \cite{WU2014,HO2019247,kosucki2020}. 

Recently, Shortcuts to Adiabaticity (STA) have been proposed as a robust alternative to optimal control of crane operations \cite{Gonzalez-Resines2017}.   
STA \cite{Guery2019} are a set of mathematical tools
to design the output of a given control operation by achieving the result of an infinitely slow evolution in shorter times.
The origin of this set of techniques is in quantum mechanics, 
but a variety of models have already shown that they are easily adaptable to 
classical mechanics \cite{Jarzynski2013, Jarzynski2017, Li2011, Faure2019, Iram2020, Stefanatos2018} 
and in particular to the control of crane operation \cite{Gonzalez-Resines2017}. 
Torrontegui et al studied the STA transport of a load with an overhead crane 
and noticed that applying the driving force in a massive mediating system (MMS), 
whose trajectory determines the evolution of the target system, but is not influenced by its backaction, 
leads to control operations that achieve the desired output regardless of the initial conditions of the system \cite{Torrontegui2017}. 
This is a desirable feature in any control operation, but
as a counterpart, it was shown that the energy consumption of the operation was dominated by the power required to change the inertia of the MMS. Thus, the robustness provided by the MMS came with a price, increased energy consumption. 
This trend was later confirmed in different setups \cite{Tobalina2018,Tobalina_2019}.  

In this letter we present a control system that reduces the consumption of the operation 
while mantaining the robustness provided by the MMS. 
We test it for crane hoisting operations designed by STA, and analyze its validity and
energetic advantage with respect to other control approaches.
Our control system uses a spring-loaded MMS connected to the target system 
(the hoisted load in our specific problem) through an energetically passive guiding system (EPGS) 
that translates the oscillating motion of the MMS into the STA trajectory of the target system. 
Our approach removes the leading consumption term from the energetic bill of the operation, 
i.e. the inertial term of the MMS \cite{Torrontegui2017}. 

\begin{figure*}[t!]
  \begin{center}
    \includegraphics[width=\linewidth]{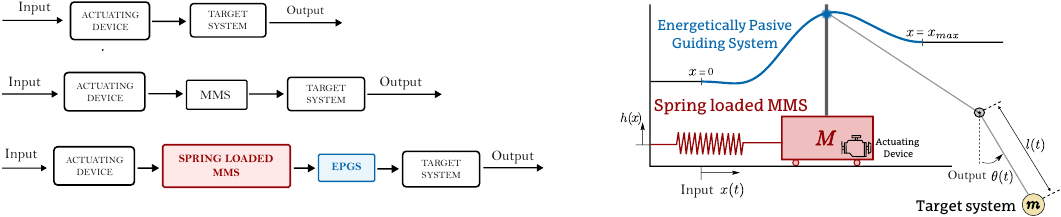}
    \caption{\label{fig:fig1}
    Left: The three conceptual control approaches studied in this letter, with the parts involved in each
    of them. Top: Direct Control; 
    Middle: Mediated Control; 
    and Bottom: Energetically Efficient Mediated Control. 
    Right: Schematic of the mechanical assembly that controls the hoist of our load $m$ (target system).
    The control system (MMS) is simply a much heavier mass ($M$) than the one we intend to lift,
    that periodically oscillates following a harmonic oscillation.
    This MMS transfers the horizontal displacement to a cart moving in a guiding rail (EPGS) above the MMS.
    The cart connects with the pulley that lifts the mass $m$,
    such that its design, i.e. its height $h(x(t))$, effectively controls the lift as designed, $l(t)$.
    }
  \end{center}
\end{figure*}

The innovation of our proposal is twofold. 
(i) It introduces an EPGS that determines the output of the control protocol,
and (ii) it features a spring-loaded MMS that recycles energy for subsequent operations. 
In the specific model analyzed here, the preasambled EPGS determines the 
trajectory of the load with no direct consumption, 
while the spring loaded troley recorvers most of the kinetik energy 
that drives the hoist as potential energy usable for lifting other loads. The implementation of our proposal for this particular operation is depicted in Fig. \ref{fig:fig1}.  
More generally, we argue that our approach establishes a new control paradigm 
that provides robustness against disturbances and reduced energy consumption, 
which can be adapted, beyond specifics of each particular scenario, to any control operation.

%
%

%
%
\paragraph*{\textbf{Model of the target system and design of STA hoist}}\label{sec:design}

A hoisting operation consists on lifting a load that hangs freely from a pulley.
The time-dependent parameter that determines the trajectory of the load is $l(t)$, the length of the rope between the pulley and itself. The hoisting process is specified by the boundary values
$l(0) = l_0$ and $l(t_f) = l_f$, where $l_0>l_f$ and $t_f$ is the duration of the operation.
Additionally, we impose
$\dot{l}(t_b) = \ddot{l}(t_b) = 0,$  for smoothness,
where $t_b$ stands for both boundary times, 0 and $t_f$.
The Lagrangian that describes the target system is
\begin{align}
  \label{Lagrangian_principal}
  \mathcal{L} = \frac{1}{2}m(\dot{l}^2+l^2\dot{\theta}^2)+mgl\cos\theta,
\end{align}
where $\theta$ is the angle of the rope with the vertical, and $g$ the gravity.
Using now the Euler-Lagrange equation,
$\frac{d}{dt}\left(\frac{\partial\mathcal{L}}{d\dot{\theta}}\right)-\left(\frac{\partial\mathcal{L}}{\partial\theta}\right) = 0$,
we get $2\dot{l}\dot{\theta} + l\ddot{\theta} + g\sin\theta = 0$.
Equivalently, we may use the horizontal displacement of the load as the variable 
to describe the motion of the target system, which, 
in the small oscillation regime, leads to a simple harmonic oscillator
$\ddot{q}+\Omega^2q = 0$, 
where $q= l sin (\theta)$ and we have defined
$\Omega^2(t) = \frac{g}{l}-\frac{\ddot{l}}{l}$.
This motion is known to have an invariant \cite{Lewis1982} of the form
$I = \frac{1}{2m}\left[ \rho p-m\dot{\rho}q  \right]^2 + \frac{1}{2}m\Omega_0^2\left(\frac{q}{\rho}\right)^2$,
%
where $p = m\frac{dq}{dt} = m\dot{q}$ is the momentum of the load and $\rho$ is an auxiliary function that satisfies the Ermakov equation
$\ddot{\rho} + \Omega^2\rho = \frac{\Omega_0^2}{\rho^3}$
with $\Omega_0$ being an arbitrary value which is typically defined as,
$\Omega_0^2 = \Omega^2(0) = g/l$  for convenience
(note that this is satisfied by the boundary condition $\ddot{l}(0) = 0$).
Following the invariant based inverse engineering theory \cite{Chen2010} of the STA,
this invariant guarantees reaching the unexcited target state so long 
as a set of boundary conditions is satisfied: 
$\rho(0) = 1; \quad \rho(t_f) = \gamma = \sqrt{\Omega_0/\Omega_f} = \sqrt[4]{l_f/l_0}; \dot{\rho}(t_b) = \ddot{\rho}(t_b) = 0.$

Due to the complex relationship between the time-dependent parameter of the target system and the auxiliary parameter in the invariant,
we ditch the usual inverse engineering strategy 
(starting with an ansatz for $\rho(t)$ and get $l(t)$ from the Ermakov equation), and instead follow a semidirect approach, 
combined with numerical optimisation \cite{Palmero2015}.
We first set an ansatz for $l(t)$ that satisfies
the boundary conditions $l(0) = l_0$, $l(t_f) = l_f$ and $\dot{l}(t_b) = \ddot{l}(t_b) = 0$,
and we leave some free parameters in this ansatz.
We then solve the Ermakov equation,
using the now analytical expression for $\Omega$.
This equation needs to be numerically solved,
so we fixed the free parameters using 
MATLAB's \emph{fminsearch} function,
which uses the Nelder-Mead algorithm \cite{Lagarias1998}
such that they minimise the boundary conditions for the $\rho$.
We borrow an energy expression from quantum mechanics
to define a cost function that considers all boundary conditions
\begin{align}
  E_{exc} = \frac{\hbar}{4\Omega_0}\left(\dot{\rho}^2+\Omega^2\rho^2+\frac{\Omega_0^2}{\rho^2}\right).
\end{align}
\paragraph*{\textbf{Control system}}\label{sec:model}

We consider three possible control systems to drive the STA hoist, see Fig. \ref{fig:fig1},
and compare them in terms of robustness and energy consumption.  
Direct Control (DC) is the most straigthforward, directly exerting the force over the pulley 
that controls the length of the rope. 
In the Mediated Control (MC), the driving force could be applied to a MMS
(of mass $M \gg m$) that is horizontally moved following the same motion as the target system,
and is direcly connected to the latter passing the rope through a pulley. 
Finally, we introduce an Energetically Efficient Mediated Control (EEMC),
where instead of manipulating the horizontal position of the MMS in time,
we will initialise it with a large amount of potential energy,
and will let it periodically oscillate following a harmonic motion
$x(t) = \frac{x_{max}}{2}(1+\cos(\omega t))$, where $x_{max}$ is the amplitude of the oscillation,
and $\omega$ the angular frequency.
This harmonic oscillation will transfer the horizontal displacement $x(t)$
to our EPGS, a massless cart moving along a rail that is assembled above the MMS.
This cart is directly connected to the target system
through a rope that pivots on the pulley.
Since the horizontal position of the cart is given by $x(t)$,
we need to control the distance between the cart and the pulley,
and effectively the hoist ($l(t)$), by modulating the height of the EPGS $h(x)$.
With simple trigonometry, we obtain
\begin{align}
  h(x) = \sqrt{[L-l(t[x])]^2-[D-x(t)]^2},
\end{align}
where $L$ is the total length of the rope,
$D$ the distance between the pulley and the cart at the maximum contraction point,
and $l(t[x])$ is the design for the lift we numerically obtained
in the previous section after changing to position parameters,
that can be deduced from the harmonic oscillation equation as
$t(x) = \frac{1}{\omega}\arccos\left(1-\frac{2x}{x_{max}}\right).$

While the DC has a simple implementation, and is the most energy efficient of all,
it heavily depends on the initial configuration of the load, 
with its subsequent drop in the fidelity of reaching the desired target state (see \cite{Torrontegui2017}).
The MC solves this dependence and provides precise operations regardless of the initial configuration of the load
at a cost of a much higher energy consumption \cite{Torrontegui2017, Tobalina2018}.
The EEMC introduced here retains the benefits of the MC, with an energy consumption 
barely larger than for the DC in the ideal case as we will see in the following. 
This comes at the cost of a more cumbersome setup and a large initialising potential energy,
but as we will argue this can be very cost effective for cyclical processes.

%
\begin{figure}[t!]
  \begin{center}
  \includegraphics[width=\linewidth]{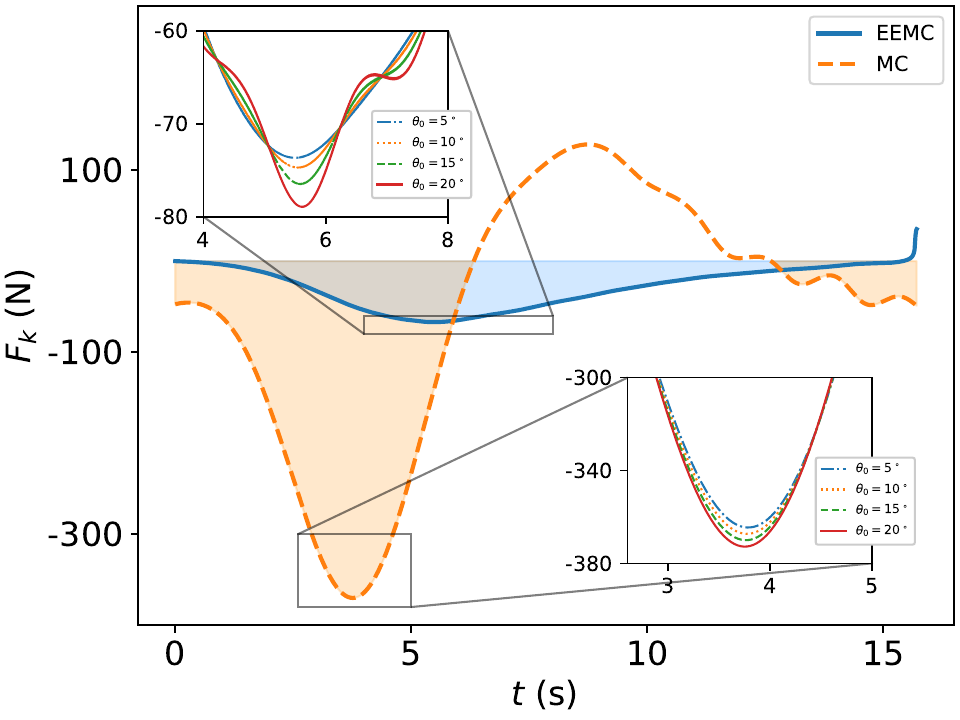}
  \caption{\label{fig:kickforce}
  Kick force $F_k$ vs time for the EEMC (solid blue)
  and the MC (dashed orange). Shaded regions indicate the forces relevant for the consumption ($\eta = 0$). Insets show the different kick forces corresponding to varying initial conditions of the load.  Parameters are: $m = 5$ kg, $M = 1000$ kg, $\omega = 0.2$ s$^{-1}$,
  $L = 20$ m, $D = 12$ m, $l_0 = 5.5$ m,  $l_f = 0.5$ m, $\mu = 0.02$, and $x_{max} = 8$ m.
  }
  \end{center}
\end{figure}
%

%
%
%
\paragraph*{\textbf{Energy consumption}}\label{sec:energy}

The amount of energy required to carry out the control operation 
amounts to the energy consumed by the actuating device to produce the force that drives it.
We assume a simplified model, 
with friction only over the MMS, 
a reasonable approximation since $M\gg m$.

We first write the Lagrangian for the target and MMS in our EEMC method:
\begin{align}
  \label{Real_Lagrangian}
  \mathcal{L}_{TS} &= \frac{1}{2}m(l'^2\dot{x}^2 + l^2\dot{\theta}^2 + mgl\cos\theta),\nonumber\\
  \mathcal{L}_{MMS} &= \frac{1}{2}M\dot{x}^2 - \frac{1}{2}M\omega^2\left(x-\frac{x_{max}}{2}\right)^2 + F_kx,
\end{align}
where we applied the chain rule such that $\dot{l} = l'\dot{x}$.
In the second line, on top of the kinetik energy and the potential energy of the control mass,
we added a term produced by some external force $F_k$ we tagged as \emph{kick force},
which will be responsible of keeping the control system following its ideal harmonic motion,
necessary to compensate for the friction and backaction.
To account for the friction, we will obtain the equations of motion
by generalising the Euler-Lagrange equation as
$\frac{d}{dt}\left(\frac{\partial\mathcal{L}}{\partial\dot{q}}\right)-\left(\frac{\partial\mathcal{L}}{\partial q}\right) + \frac{\partial \mathcal{F}}{\partial \dot{q}}= 0$,
where $\mathcal{F} = \mu \dot{x}^2/2$ is Rayleigh's dissipation function \cite{Goldstein2002}.
Following the generalised Euler-Lagrange equation,
from Eq. \eqref{Real_Lagrangian}
we now get the set of equations of motion that describe the dynamics:
\begin{align}
  0 &= 2l'\dot{x}\dot{\theta} + l\ddot{\theta} + g\sin\theta \label{evolutiontheta}\\
  F_k &= ml'\left(l'\ddot{x}+l''\dot{x}^2-l\dot{\theta}^2-g\cos\theta\right)\nonumber\\
  &+M\ddot{x}+M\omega^2\left(x-\frac{x_{max}}{2}\right)+\mu \dot{x}.\label{evolutionxtotal}
\end{align}
Once $F_k$ is known,
we can define the power exerted 
as $P = F_k\dot{x}$.
Following discussions in \cite{Torrontegui2017, Tobalina2018}
we can define the energy consumption of the kick force as
\begin{align}
  \label{consumption}
  \mathcal{E} = \int\limits_0^{t_f} P_+dt + \eta\int\limits_0^{t_f}P_-dt,
\end{align}
where we now distinguish the positive powers $P_+$
for times where the kick force is pushing in the same sense as the motion,
and the negative power $P_-$
when the kick force pushes against the motion.
The factor $\eta$ can in theory be anything between -1 and 1.
A value of 0 would mean an ideal braking system,
and positive value would imply some sort of kinetic energy recycling system.
Here, we will stick simply to the case $\eta = 0$, 
although a small negative $\eta$ would be more realistic, specially for strong braking forces.

For the MC case, we set the height of the EPGS as $h = 0$ for all times,
so $l(t) = x(t)$.
In Eqs. \eqref{evolutiontheta} and \eqref{evolutionxtotal} we only need to replace
$l' = 1$ and $l'' = 0$, and after obtaining the new $F_k$ we will calculate the benchmark
energy consumption using Eq. \eqref{consumption}, with the same condition $\eta=0$ as before.

%
\begin{figure}[t]
  \begin{center}
  \includegraphics[width=\linewidth]{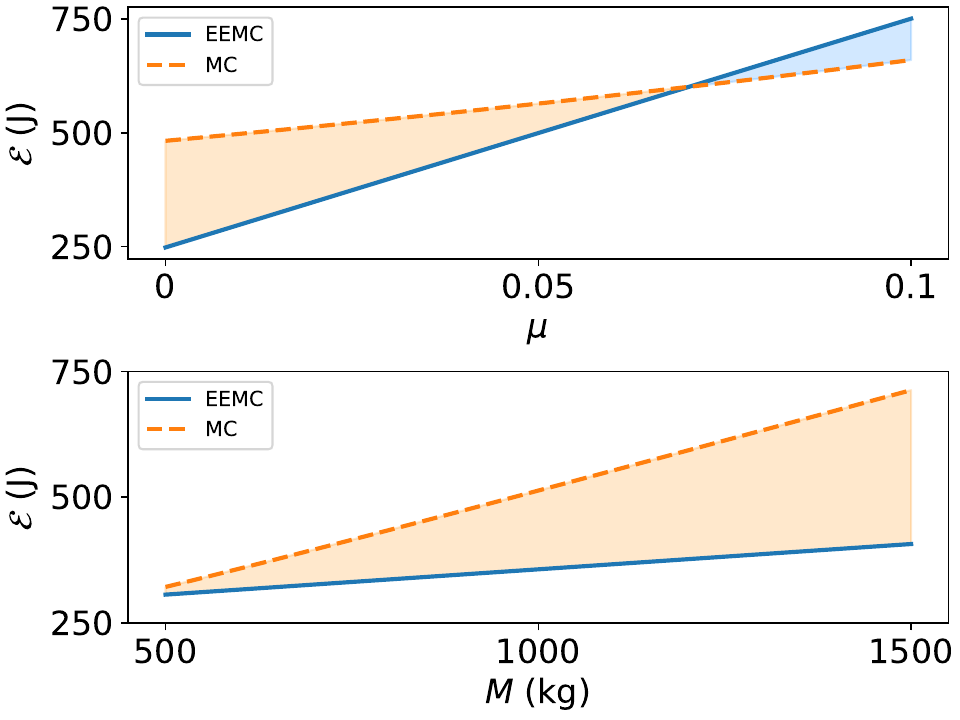}
  \caption{\label{fig:Evsmu}
  Upper panel: Consumption energy for the EEMC (blue solid) 
  and the MC (orange dashed) for different values of friction.
  Lower panel: Consumption energy for the EEMC (blue solid) 
  and the MC (orange dashed) for different values of the MMS. Orange shaded regions indcate favorable regimes for the EEMC.
  Except for $\mu$ in the upper panel and $M$ in the lower panel, the rest of the parameters are the same as in Fig. \ref{fig:kickforce}.
  }
  \end{center}
\end{figure}
%

%
%
%
\paragraph*{\textbf{Results}}\label{sec:results}

Figure \ref{fig:kickforce} compares the kick forces of EEMC and MC in a benchmark operation.
This example shows that the EEMC requires much smaller forces, therefore, it has a huge energetic advantage per cycle.
Moreover, braking forces (positive forces in our case) only appear briefly at the end
of the driving in the guided case, whereas the unguided benchmark case has notable braking forces.
These do not affect in our annalysis since we set $\eta = 0$, see Eq. \eqref{consumption}, 
but a braking force may also have some sort of comsumption,
a further argument in favour of our approach.
Notably, Fig. \ref{fig:kickforce} shows that the kick force required to drive the operation with the EEMC 
barely changes for substantially different initial orientatios of the load (comparable to the MC case), 
showing its robustness agains changing initial conditions, 
retaining the benefits of the MC over the DC. 

Figure \ref{fig:Evsmu} shows the energy consumed with the MC and the EEMC for varying friction coefficients. It reveals a reduced consumption of the EEMC for low friction setups. As friction increases, the advantage drops, until the situation is even reverted.
The reasoning behind this is that, with the more complex ensemble,
the kick force in Eq. \eqref{evolutionxtotal} has two leading terms ($\sim M$).
Near the ideal case, these two terms practically offset each other,
but as we get away from the ideal driving, both terms will start to contribute positively. Anyhow, mechanical engineering companies report friction coefficients ranging from 0.001 to 0.005 with standard methods (even lower values are achievable through more advanced mechanisms such as cylindrical roller bearings) \cite{NTN2016}, values well within the regime where the EEMC yields reduced energy consumption.  In the lower panel, Fig. \ref{fig:Evsmu} shows how the comsumption scales
as the mass of the MMS increases (recall that heavier MMS imply increased robutness of the operation agints external perturbations). 
Here the tendency is clearly favorable for the EEMC. 

Table \ref{tab:table1} shows some example values comparing consumptions with EEMC and MC in the context of the minimum energy required 
for the lift\footnote{
  The minimum energy is the potential energy difference between the initial and 
  final position. In general this corresponds to the energy consumption of the DC case,
  although for very short hoist times $l(t)$ might be nonmonotonous, increasing the
  energy consumption in setups with $\eta\neq 1$.
}, 
and indicates in each case the amount of cycles neccesary to overcome the initial potential energy loaded in the spring, a cost not present in the MC. 
In the ideal case, i.e., no friction, the EEMC only requires an excess 1.2\%
energy to drive the operation, 
whereas the benchmark MC requires almost double that energy.
For cases with a moderate amount of friction, 
usually a handful cycles is enough to compensate for the potential energy
we initialise the harmonic oscillator with. 
As a caveat, the EPGS must imply some sort of additional expenditure 
in manufacturing, installation, or maintaninance. 
We regard these costs as independent from the control operation itself 
and do not include them in our analysis. 
In any case, they would also be recouped after a certain number of cycles.

\begin{table}[t]
  \caption{\label{tab:table1}%
  Values of consumption energy and the consumption energy required in 
  excess of the strictly minimum potential energy ($E_p = mg\Delta l = 245$ J)
  for some of the values in Fig. \ref{fig:Evsmu}. The last column shows how
  many cycles it would take to compensate for the initial potential energy requiered in the EEMC ($E_k = \frac{1}{2}M\omega^2(x_{max}/2)^2 = 320$ J).
  }
  \begin{ruledtabular}
    \begin{tabular}{llccc}
      \textrm{$\mu$}&
      \textrm{case}&
      \textrm{$\mathcal{E}$ (J)}&
      \textrm{$\frac{\mathcal{E}-\Delta E_p}{\Delta E_p}$ (\%)}&
      \textrm{\# cycles}\\
      \colrule
      0 & EEMC & 247.9 & 1.2 & 2\\
        & MC & 482.2 & 96.8 & \\
        \hline
      0.02 & EEMC & 349.0 & 42.2 & 2\\
      & MC & 513.4 & 109.6 & \\
        \hline
      0.04 & EEMC & 449.0 & 83.3 & 3\\
      & MC & 546.6 & 123.1 & \\
        \hline
      0.06 & EEMC & 549.5 & 124.3 & 8\\
      & MC & 581.9 & 137.5 & \\
        \hline
      0.08 & EEMC & 650.1 & 165.3 & -\\
      & MC & 619.6 & 152.9 & \\
    \end{tabular}
  \end{ruledtabular}
\end{table}

\paragraph*{\textbf{Conclusions}}\label{sec:conclusions}

We have presented the EEMC for a hoisting operation; a mechanical control system that involves an spring loaded MMS and an EPGS. 
The MMS provides robustness against changes in the initial conditions of the load, while the EPGS converts efficiently the harmonic motion of the oscillating MMS
into the specific motion we need for lifting a hanging mass swiftly and without excitations.
The proposed control system requires an initial input of potential energy, however,
it yields reduced energy consumption in a single operation, 
quickly reaching cost-efficiency in cyclical operations.

The consumption of EEMC scales worse with friction compared with that of the MC.
However, it shows a clear energetical advantage for a realistic range of friction coefficients.
Since we present here a rather simple and linear mechanical system,
seeking efficiency through low friction coefficients is something desirable,
that will not necessarily be detrimental to our control capabilities.
This is something achievable by lubricating the rails \cite{Stock2011, Lu2012, Lundberg2015}.
On the other side, the proposed method scales much better against higher MMS masses $M$.
We use large control MMS,
since this mass is the one giving stability to the target system,
reducing the effect of initial conditions and backaction from the primary system. 

The method of choice to design the control ouput,
known as `Invariant based inverse engineering',
offers ample flexibility.
On top of the conditions that guarantee a fast
and excitationless driving of the target system,
we could add additional conditions to simultaneously optimise, 
e.g., minimising the consumption energy,
minimising the usage of material for the EPGS,
reducing the driving to the shortest possible time,
or making the driving more robust with respect to some noise, or wind. 
Moreover, the control system introduced here could be used to produce 
control outputs designed with other techniques such as optimal control theory.

Finally, we would like to highlight the potential importance of this method to other disciplines. 
Besides being directly applicable to any other mechanical control operation, 
the same road that led from quantum mechanics to classical mechanics can lead the way back. 
In fact, energy efficiency is also a major concern in quantum technologies \cite{Auffeves2022}.
The idea of the energetically passive system that transforms a predetermined motion into a specific control output
could be adopted in the context of quantum thermodynamics \cite{Kosloff2013}, for example, to drive a single ion in a Paul trap undergoing an Otto cycle \cite{Rosnagel2016}.

{\bf Acknowledgments}:

We are greatful for funding to the Spanish Ministry of Science and Universities
through projects PID-2021-125823NA-I00 and PID-2021-126277NB-I00,
the Basque Government through project IT1470-22 and KUBIBIT project of Elkartek program,
and to the EU Flagship on Quantum Technologies through OpenSuperQ+100 (Grant No. 101113946).
This project has also received support from the Spanish Ministry for Digital Transformation 
and of Civil Service of the Spanish Government through the QUANTUM ENIA project call - Quantum Spain.
MP is also thankful to Dario Poletti and Bo Xing for hosting him at 
the Singapore University of Technology and Design, where this 
project was partially developed.

\bibliography{Bibliography}{}

%
%
%

%
%
\end{document}